\documentclass[12pt,epsf]{article}
\usepackage{graphicx, amsmath}

\begin{document}
\sloppy
\begin{flushright}{SIT-HEP/TM-56}
\end{flushright}
\vskip 1.5 truecm
\centerline{\large{\bf Remote Inflation as hybrid-like}}
\centerline{\large{\bf sneutrino/MSSM inflation}}
\vskip .75 truecm
\centerline{\bf Tomohiro Matsuda\footnote{matsuda@sit.ac.jp}}
\vskip .4 truecm
\centerline {\it Laboratory of Physics, Saitama Institute of Technology,}
\centerline {\it Fusaiji, Okabe-machi, Saitama 369-0293, 
Japan}
\vskip 1. truecm

\makeatletter
\@addtoreset{equation}{section}
\def\theequation{\thesection.\arabic{equation}}
\makeatother
\vskip 1. truecm

\begin{abstract}
\hspace*{\parindent}
A new scenario of hybrid-like inflation 
is considered for sneutrino and MSSM fields.
Contrary to the usual hybrid inflation model, the direct coupling
 between a trigger field and the sneutrino/MSSM inflaton field is
 not necessary for the scenario. 
The dissipation and the radiation from the sneutrino/MSSM
 inflaton can be written explicitly by using the Yukawa couplings.
Remote inflation does not require the shift symmetry or cancellation
in solving the $\eta$ problem.
\end{abstract}

\newpage
\section{Introduction}
According to the recent cosmological observations, inflation 
has become the
paradigm for early cosmology and appears to be the most
successful cosmological model giving the primordial large scale
structure of the Universe. 
Considering typical slow-roll inflation, there are at least two
scenarios for the velocity damping: the original (supercooled)
inflation and warm inflation in which a strong
diffusion produces significant friction for the inflaton
motion\cite{warm-inflation-original}.  
Recently, hybrid-like extension of warm inflation has been proposed
in Ref.\cite{remote-matsuda}, in which a new possibility is shown that 
the radiation produced concurrently during warm inflation may keep
symmetry restoration in a remote sector.
As a consequence, the
false-vacuum energy of the remote sector may dominate the energy density
during inflation. 
The situation is similar to thermal inflation, however in contrast to
the standard thermal inflation model, the radiation in remote inflation
is sourced by the dissipation of the inflaton field.
Remote inflation is a new inflationary model based on thermal
inflation\cite{thermal-inflation},  hybrid
inflation\cite{hybrid-inflation}  and warm
inflation\cite{warm-inflation-original} and is expected to solve
problems in these classic models.
In this paper we consider remote inflation for sneutrino
inflation - inflation caused by the scalar
supersymmetric partner of a heavy singlet
neutrino\cite{sneutrino-inflation, Mazumdar-reviewer} - and MSSM
inflation - inflation 
caused by the fields in the minimal supersymmetric standard model
(MSSM)\cite{MSSM-inflation-0, MSSM-inflation-matsuda}.

Remote inflation is based on warm inflation.
Warm inflation for a sneutrino field has been discussed in
Ref.\cite{sneutrino-warm}, in which monomial potentials
constructed with only the right-handed sneutrino field realize
warm inflationary regime for chaotic inflation.
Since the conditions for warm inflation and supercooled inflation are
exclusive, the parameter space for successful sneutrino inflation must
be complemented by considering warm inflation.
With regard to the problems and the benefits of sneutrino/MSSM, hybrid
and warm inflation scenarios, remote inflation may give a very attractive
situation in which hybrid-like inflation is realized in the warm
inflationary regime without introducing additional direct 
coupling between the inflaton and a trigger field.
Applying remote inflation to sneutrino and MSSM fields, it may be
possible to find a successful scenario in which the baryon asymmetry of
the Universe is created by the non-equilibrium decay of the 
inflaton.\footnote{See Fig.1.}
\begin{figure}[h]
 \begin{center}
\begin{picture}(320,180)(0,270)
\resizebox{12cm}{!}{\includegraphics{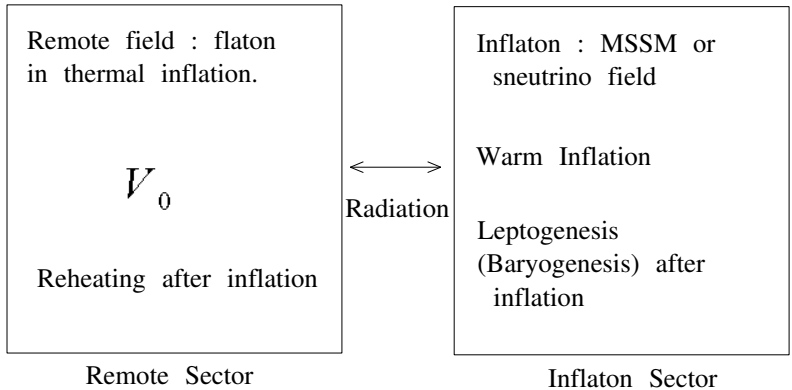}} 
\end{picture}
\label{fig:eps}
 \caption{Remote inflation resembles hybrid inflation. Remote field
  plays the role of the trigger field in hybrid inflation, which causes
  reheating the Universe. The explicit interaction in the conventional
  hybrid potential is replaced by the thermal background. Also, the
  remote sector resembles thermal inflation. The source of the radiation
  is the dissipation in the inflaton sector, where warm inflation occurs.}  \end{center}
\end{figure}

The idea of remote inflation is very simple.
Considering the standard hybrid inflation model, the typical form of
the hybrid-type potential is given by 
\begin{equation}
V(\phi,\sigma)=V(\phi) + \frac{g^2}{2} \sigma^2\phi^2
+\frac{\lambda}{4}\left(\sigma^2-M^2\right)^2,
\end{equation}
where the vacuum energy during inflation is $V_0\simeq \lambda M^4/4$.
The trigger field $\sigma$ stays in its false-vacuum state 
defined by $\sigma=0$ for $\phi>\phi_c$.
Here the critical point that defines the end of inflation is given by
\begin{equation}
\phi_c\equiv \sqrt{\frac{\lambda}{g}}M.
\end{equation}
Contrary to the usual hybrid inflation scenario, either
normal, inverted \cite{inverted-hybrid}
brane-motivated \cite{brane-inflation-original, matsuda-non-tachonic,
matsuda-topological-brane},
or thermal \cite{matsuda-thermal-brane},
remote inflation does not have direct couplings between inflaton and
trigger fields. 
The trigger fields $\sigma_i$ in the remote inflation model are
placed in a remote sector and stay in their false vacuum states during
inflation due to the thermal effects created by the diffusion of the
inflaton field.
Namely, the force that keeps $\sigma_i$ at their false-vacuum state is
not the 
effective mass $m_\sigma^2 \sim g^2 \phi^2$ 
caused by the direct coupling between the inflaton and
the trigger field, but the thermal effect of the radiation concurrently
produced during warm inflation. 

Remote inflation requires symmetry restoration in the remote sector.
To illustrate some typical features of finite temperature effects in the
remote sector, here we consider a real scalar field and a potential: 
\begin{eqnarray}
\label{ini-the}
{\cal L}&=&\frac{1}{2}\partial_{\mu}\sigma\partial^{\mu}\sigma-V(\sigma)\nonumber\\
V(\sigma)&=&V_0-\frac{1}{2}m_\sigma^2\sigma^2+\frac{1}{4}\lambda\sigma^4,
\end{eqnarray}
where $V_0$ is tuned so that the cosmological constant vanishes at the
true minimum. 
The phenomenon of high-temperature symmetry restoration
can be understood by the finite-temperature effective potential given by
\cite{EU-book-kolb}
\begin{equation}
V_T(\sigma_c)=V(\sigma_c)+\frac{T^4}{2\pi^2}\int^{\infty}_{0}
dx \ln \left[1-\exp\left(
-\sqrt{x^2+\frac{-m_\sigma^2+3\lambda\sigma_c^2}{T^2}}
\right)
\right],
\end{equation}
where $V(\sigma_c)$ is the one-loop potential for zero-temperature with
the classical field $\sigma_c$:
\begin{equation}
V(\sigma_c)=-\frac{1}{2}m_\sigma^2\sigma_c^2+\frac{1}{4}\lambda\sigma_c^4
+\frac{1}{64\pi^2}\left(-m_\sigma^2+3\lambda\sigma_c^2\right)^2
\ln \left(\frac{-m_\sigma^2+3\lambda\sigma_c^2}{\mu^2}\right),
\end{equation}
where $\mu$ is a renormalization mass scale.
At high temperatures, $V_T$ can be expanded near the symmetric point
($\sigma_c=0$) as
\begin{equation}
V_T\simeq V(\sigma_c)+\frac{1}{8}\lambda T^2 \sigma_c^2+ {\cal O}(T^4),
\end{equation}
which suggests that the temperature-corrected effective mass at 
$\sigma_c=0$ changes sign at the critical temperature given by
\begin{equation}
T_c \simeq \frac{2m_\sigma}{\lambda^{1/2}}.
\end{equation}
More generally, a background thermal bath which couples
to $\sigma$ can be introduced.
Then we obtain a typical form of the potential with a thermal correction
term, which is given by
\begin{equation}
V=V_0+\left(g^2 T^2 -\frac{1}{2}m_\sigma^2\right)\sigma^2+ ...,
\end{equation}
where $g$ denotes the effective coupling of $\sigma$ to the fields in the
thermal bath. 
In this case, the critical temperature is given by
\begin{equation}
T_c \simeq \frac{m_\sigma}{2g}.
\end{equation}
In this paper, we assume that the couplings to the background fields
are significant and $T_c$ is  given by
$T_c \simeq \frac{m_\sigma}{2g}\sim m_\sigma$. 
Here $\lambda \ll 1$ is possible for a flat potential in supersymmetric
theory, which gives typically the false-vacuum energy 
$V_0 \sim m_\sigma^2 M^2$ for the quartic potential, 
where $M^2\sim
m_\sigma^2/\lambda \gg m_\sigma^2$  denotes the vacuum
expectation value of the field $\sigma$ in the true vacuum.

\section{Remote inflation with dissipative sneutrino}
The interaction of the sneutrino inflaton field with other fields
is very important in sneutrino inflation, since
Yukawa couplings are always needed for the seesaw
mechanism\cite{see-saw} that gives the tiny masses to the light
neutrinos.
The heavy singlet neutrinos usually have masses $10^{10}$ to $10^{15}$
GeV, but one may consider a smaller mass for a singlet
neutrino. 
For our scenario the simplest and sufficient model is only one sneutrino
acting as the inflaton whose dissipation sources radiation during
inflation.
The interaction relevant for the model is extracted from the
supersymmetric version of the seesaw model as
\begin{equation}
{\cal L}_I\sim -|h_N|^2 |\phi_N|^2 |H_u|^2 + h_t H_u \bar{t}_R t_L
+h.c.,
\end{equation}
where $\phi_N$ is the sneutrino inflaton, $H_u$ is the Higgs doublet
giving masses to the up quarks and  $t_R, t_L$ are the top quarks.
Dissipative effects are caused by the coherent excitation of the $H_u$
field, which then decays into fermions (standard-model quarks) 
with the decay rate
\begin{equation}
\Gamma_t =\frac{3h_t^2}{16\pi}m_{H_u}\sim \frac{3h_t^2}{16\pi}h_N \phi_N.
\end{equation}
The condition for the thermalization of the decay product is given by
\begin{equation}
\Gamma_t >H,
\end{equation}
which puts a natural upper bound for the energy scale of warm
inflation. 
Following Ref.\cite{warm-inflation-original} and \cite{sneutrino-warm},
the typical form of the dissipative coefficient $\Gamma$ for warm
inflation is given by 
\begin{equation}
\Gamma \simeq \frac{\sqrt{\pi}}{20}\frac{h_N^2 h_t^2}{(4\pi)^2}
\phi_N\equiv A_N \phi_N.
\end{equation}
The strength of the thermal damping is measured by the rate 
$r_\Gamma$ given by 
\begin{equation}
r_\Gamma\equiv \frac{\Gamma}{3H},
\end{equation}
which can be used to rewrite the field equation as
\begin{equation}
\ddot{\phi}+3H(1+r_\Gamma)\dot{\phi_N}+V(\phi_N,T)_{\phi_N}=0,
\end{equation}
where the subscript denotes the derivative with respect to the field.
Considering a typical quadratic potential for the singlet neutrino,
a simple form can be used for the potential;
\begin{equation}
V(\phi_N,T)_{\phi_N}\simeq M_N^2 \phi_N,
\end{equation}
where $M_N$ is the mass of the sneutrino.
From the above equations the slow-roll velocity of the sneutrino
inflaton for the strongly dissipating regime ($r_\Gamma >1$)
is given by
\begin{equation}
\dot{\phi}_N \simeq \frac{M_N^2 \phi_N}{\Gamma}\simeq\frac{M_N^2}{A_N},
\end{equation}
which does not depend on the temperature.
Note that the slow-roll velocity is independent of the temperature of
the Universe. 
The conventional slow-roll parameters are given by
\begin{eqnarray}
\epsilon &\equiv& \frac{M_p^2}{2}\left(\frac{V_\phi}{V}\right)^2 
<(1+r_\Gamma)\nonumber\\
\eta &\equiv& M_p^2\left(\frac{V_{\phi\phi}}{V}\right)
<(1+r_\Gamma).
\end{eqnarray}
In addition to the standard slow-roll conditions for the inflaton motion, 
slow-roll approximation for warm inflation requires slow-variation of
$\Gamma$, which leads to an additional slow-roll parameter 
\begin{equation}
\beta \equiv M_p^2 \left(\frac{\Gamma_{\phi_N}V_{\phi_N}}{\Gamma V}\right)
\simeq \frac{M_N^2}{H^2}\sim \eta <(1+r_\Gamma).
\end{equation}
This equation gives a trivial condition for the present model.
Similarly, $T$-dependence of $\Gamma$ leads to a condition for the
radiation\cite{warm-inflation-original}
\begin{equation}
c\equiv \left|\frac{d\ln \Gamma}{d\ln T}\right| =
 \left|\frac{d\ln \Gamma}{d\ln \phi_N}\frac{d\ln\phi_N}{d\ln T}\right| 
=4,
\end{equation}
which does not satisfy the required condition $c<4$.
Here $c=4$ is the direct consequence of the quadratic potential,
namely $V\propto \phi_N^n$ with $n=2$.
However, in more general situations the effective potential may have 
other higher terms with $n>2$, although they may be small
compared with the quadratic term. 
Therefore, the effective $n$ would be $n=2+\epsilon_N>2$, where 
$\epsilon_N\ll 1$. 
A small shift from $n=2$ is trivial and it can be disregarded in other
scenarios, but in hybrid-type warm inflation the shift is
crucial for warm inflation\cite{consistency-warm}. 
Moreover, if initially the scalar field gets a value as large as
the Planck (or the fundamental) scale, the potential at that time may
have significant higher corrections with large $n$.
Therefore, in this paper we consider the quadratic potential, 
but the effective value of $n$ is slightly larger than
$n=2$. 
Other possibilities, the potential characterized by higher terms 
with $n>2$, or a flat potential characterized by $\sim \log \phi$
are discussed in section \ref{mssm-inf}.

During inflation, the energy conservation of the radiation energy
density gives the equation
\begin{equation}
\dot{\rho}_r +4H\rho_r =\Gamma \dot{\phi}^2,
\end{equation}
which leads to the temperature during inflation given by
\begin{equation}
T^4 = \frac{45}{2\pi^2 g_*}r_\Gamma \dot{\phi^2},
\end{equation}
where $g_*\sim {\cal O}(10^2)$ is the effective number of the light
degrees of freedom at the temperature.
A useful expression can be found from the above equation, which is
given by 
\begin{equation}
T^4 \simeq M_N^4 \frac{\phi_N}{A_N H},
\end{equation}
or equivalently;
\begin{equation}
r_T^4 \equiv \left(\frac{T}{H}\right)^4 \simeq \eta^2 
\frac{\phi_N}{A_N H}.
\end{equation}
Since the inflationary expansion is not viable when the radiation
dominates the Universe, we also consider the obvious condition for the
radiation density $\rho_r$, which is given by
\begin{equation}
\rho_r < V_0.
\end{equation}
In contrast with standard sneutrino-warm inflation\cite{sneutrino-warm}, 
in which the above 
condition determines the end of inflation, 
this equation leads to the mild condition for remote inflation;
\begin{equation}
\phi_N < \phi_N^{(r)}\equiv A_N H \left(\frac{V_0}{M_N^4}\right).
\end{equation}
If the inflaton reaches the breaking-point of the slow-roll
condition ($\phi_N^{(e)}$) ahead of the critical temperature ($T=T_c$), 
the temperature at the end of inflation ($T=T_e$) is still higher than  
the critical temperature ($T_e>T_c$).
The number of e-foldings is given by
\begin{equation}
\label{e-fol-eq}
N_e \simeq -\int^{\phi_{N}^{(i)}}_{\phi_{N}^{(e)}}
\frac{3H^2}{\dot{\phi}}d\phi +N_{th}
\simeq \frac{1+r_\Gamma}{\eta}\ln\frac{\phi_{N}^{(i)}}{\phi_{N}^{(e)}}
+N_{th},
\end{equation}
where the additional expansion of the Universe, denoted by
$N_{th}$,  is caused by thermal
inflation that starts at $T=T_e$ and ends at $T=T_c$.
Thermal inflation is defined for an inflationary expansion due to the
vacuum energy of a false-vacuum state, but in contrast to remote
inflation there is no significant source that supplies radiation during
this epoch.\footnote{Breaking of the slow-roll condition does not always
mean the end of dissipation. Here we assumed for simplicity 
that $\Gamma\propto \phi_N$ decreases rapidly after the critical point
at $\phi_N=\phi_N^{(e)}$, and as a consequence, the dissipation caused
by the inflaton oscillation is small during this period.
Numerical studies are required in order to understand more
details of the final stage of remote inflation, where oscillating
inflaton may source radiation by dissipation, evapolation or decay.} 
The e-foldings of the expansion due to thermal inflation is given by
$N_{th}\sim 
\ln(T_e/T_c)$.
On the other hand, if inflation ends when the temperature reaches
$T=T_c$ ahead of $\phi=\phi_N^{(e)}$, symmetry breaking in the remote
sector occurs just at the end of remote inflation.
In the latter case there is no thermal inflation
after remote inflation \cite{remote-matsuda}.

\subsection{Slow-roll conditions and the sources of the cosmological perturbations}
In standard warm inflation, the effective potential depends on the
temperature $T$, which can be expressed as $V(\phi_N,T)$. 
The damping rate $\Gamma$ of the inflaton field may also depend on $T$
and $\phi_i$.  
Therefore, the trajectory of warm inflation
is generally given by the inflaton and the temperature.
Contrary to the standard warm inflation scenario, $\Gamma$ in sneutrino
inflation is independent of $T$.
We thus find immediately that the damping of the inflaton $\phi_N$ is
determined independent of the radiation.
In contrast to the above argument, which suggests that the temperature
is not important for the inflaton dynamics of dissipating (not
necessarily warm)
sneutrino inflation, the temperature is important for determining
the end of inflation.
As will be shown in this section, in sneutrino-remote inflation the
phase transition in the remote sector 
triggers reheating and determines the end of inflation.
Therefore, there are two sources of the curvature perturbations
in remote inflation;
the curvature perturbations created near the horizon exit and the one
created at the inhomogeneous phase transition.
In this paper we consider these two sources and compare the magnitude of
the curvature perturbations.\footnote{In addition to these sources,
inhomogeneities of the diffusion rate $\delta \Gamma$ may be important
and it may lead to a
significant non-gaussian spectrum\cite{matsuda-warm}. 
$\delta \Gamma$ causes inhomogeneities of the inflaton velocity, which
leads to inhomogeneities of the number of
e-foldings\cite{Modulated-matsuda}. 
Namely, considering the explicit form of the number of e-foldings given
by Eq. (\ref{e-fol-eq}), the perturbations $\delta N$\cite{Delta-N-ini}
 caused by $\delta
\Gamma$ is given by
\begin{equation}
\delta N \sim N_e \frac{\delta \Gamma}{\Gamma}\sim  
N_e \left(\frac{\delta A_N}{A_N} +\frac{\delta \phi_N}{\phi_N}\right),
\end{equation}
which may be as large as the conventional curvature perturbations.
See Ref.\cite{matsuda-warm} for more details on this topic.}
Although the conditions for remote inflation are highly model-dependent,
we can find simple result at least for sneutrino inflation.
Here the slow-roll conditions are defined by $\epsilon<
r_\Gamma$(SDWI) and $\epsilon<1$(WDWI),
which determines $\phi_N^{(e)}$ as
\begin{eqnarray}
\phi_N<\phi^{(e)}_N&\simeq& A_N \frac{H^3M_p^2}{M_N^4}
\sim A_N M_p\times\left(\frac{M_p}{M_N}\right)\eta^{-3/2}\,\,\,\, (SDWI)\\
\phi_N<\phi^{(e)}_N&\simeq& M_p \eta^{-1} \,\,\,\, (WDWI).
\end{eqnarray}
Sneutrino-remote inflation uses SDWI conditions.
Note that the slow-roll condition is violated in the {\bf outer region}
 of the field space, which becomes trivial for $\eta <1$.
Considering the critical temperature $T_c$, remote inflation is possible
 for the region given by 
\begin{equation}
\phi > \phi^{(c)}_N \simeq A_N\frac{H T_c^4}{M_N^4}
\simeq A_N\frac{H}{\eta^2}\left(\frac{T_c}{H}\right)^4 \simeq
A_N\eta^{-2}r_{T_c}^3 \times T_c,
\end{equation}
where $r_{T_c}$ denotes the ratio $r_T\equiv T/H$ at the critical
temperature.
Considering the ratio between $\phi^{(e)}_N$ and $\phi^{(c)}_N$, we find
\begin{equation}
R_{e/c}\equiv \frac{\phi^{(e)}_N}{\phi^{(c)}_N}\simeq 
\frac{H^2 M_p^2}{T_c^4}\sim \frac{V_0}{T_c^4}\gg 1.
\end{equation}
The equation shows that for sneutrino inflaton field, remote inflation is
 possible during $\phi^{(c)}_N<\phi< \phi^{(e)}_N$.
Namely, we find in this model the phase transition in the remote sector
determines the end of inflation.\footnote{In contrast with remote
 inflation, the end of standard sneutrino-warm inflation is determined
 by the condition $\rho_r<V\sim M_N^2\phi_N^2$\cite{sneutrino-warm}.}

The spectrum of the cosmological perturbations created during
inflation has been expected to be scale-invariant and Gaussian, but
recent observations may suggest small anomalies.
A small departure from exact scale-invariance and a certain non-Gaussian
character \cite{Bartolo-text, NG-obs} can help reveal the 
dynamics of the inflation model.
In particular, a tiny shift in the spectrum index $n-1\ne 0$ is a
useful example\cite{EU-book} of scale invariance violation.
The spectral index is used to determine the form of the inflaton
potential. 
Observation of non-Gaussian character in the spectrum may lead to a
significant bound for the mechanism of generating curvature
perturbations.
In order to explain how these anomalies are explained, many models of
inflation and the generating mechanism of the  
curvature perturbations are proposed, in which the time when the
perturbations are generated is very important.
With regard to the strongly dissipating warm inflation, the curvature
perturbations generated at the horizon exit are given by
\begin{eqnarray}
{\cal R}_{warm}^{(ini)}=-H\frac{\delta q}{\rho+P}=
H\frac{\dot{\phi}\delta \phi}
{\dot{\phi}^2+Ts}
\simeq H\frac{\dot{\phi} \delta \phi}
{Ts},
\end{eqnarray}
which is not consistent with the expectation that the number of
e-foldings in warm inflation is determined exclusively by the inflaton
field $\phi$.
If the expectation is correct, the perturbation $\delta N$ should be
given by $\sim H \delta \phi/\dot{\phi}$. 
The solution to this problem has been discussed in \cite{matsuda-warm}.
Considering evolution during inflation\cite{matsuda-warm} based on the
$\delta N$ calculation given in Ref.\cite{Modulated-matsuda, Delta-N-ini},
the correction from the evolution is found to be given by
\begin{eqnarray}
{\cal R}&\sim & H\frac{\delta \phi}{\dot{\phi}},
\end{eqnarray}
which is consistent with the expectation.
Creation (evolution) of the curvature perturbations
during inflation is
considered by many authors\cite{Modulated-matsuda, Kofman-modulated,
A-NEW, matsuda-stop-index, roulette-inflation}.\footnote{
Here the ``evolution'' should be distinguished from the trivial
$k$-dependence of ${\cal R}^{(ini)}$.}
Besides the standard scenario in which the curvature perturbations
are created at the horizon exit or during inflation, there is an
alternative in which the curvature
perturbations are created at the end of inflation\cite{End-Modulated,
End-multi, End-multi-mat0, End-multi-mat}.
There are many other possibilities for the creation of the curvature
perturbations.
For example, creation of the curvature perturbations is possible
at preheating\cite{MSSM-inflation-matsuda, IH-PR,
IH-PR-low, kin-NG-matsuda, IH-string}, reheating\cite{IH-R, Preheat-ng},
or even at much later than the reheating period\cite{IH-pt,
curvaton-paper, curvaton-dynamics, matsuda_curvaton}.
They may be important for remote inflation, 
however for simplicity most of these alternatives are not discussed in
this paper. 

Below we first consider the creation of the curvature perturbations
at the beginning of inflation, then
we examine the inhomogeneous phase transition.
Inhomogeneities of the diffusion rate $\delta \Gamma$ may lead to
a significant creation of the curvature perturbations,
however for the simplicity of the argument, they 
are not considered in this paper.
The baryon number of the Universe caused by the decay of the sneutrino
inflaton\cite{sneutrino-inflation, Mazumdar-reviewer} (leptogenesis) is discussed in section \ref{lepto-sn}.

Due to the thermal effect, the amplitude of the inflaton perturbation
is enhanced during strongly dissipating warm inflation. 
From the Langevin equation, the root-mean square fluctuation amplitude
of the inflaton field $\delta \phi$ after the freeze out 
is obtained to be\cite{warm-inflation-original}
\begin{eqnarray}
\label{pert-phi}
\delta \phi_{\Gamma >H}&\sim& 
(\Gamma H)^{1/4}T^{1/2}
\sim r_\Gamma^{1/4}r_T^{1/2}H
\end{eqnarray}
where $r_T$ denotes the ratio between $T$ and $H$, defined by
 $r_T\equiv T/H$, which is assumed to be $r_T >1$ due to the definition
of warm inflation.
We find for the sneutrino-remote scenario characterized by 
$\Gamma = A_N \phi_N$ and $V(\phi_N)\sim V_0 + M_N^2 \phi_N^2/2$;
\begin{eqnarray}
r_T&\simeq& \frac{M_N \phi_N^{1/4}}{A_N^{1/4}H^{5/4}}\sim \eta^{1/2} 
\left(\frac{\phi_N}{A_NH}\right)^{1/4}\\
r_\Gamma &\simeq& \frac{A_N\phi_N}{3H}.
\end{eqnarray}
Therefore, for SDWI, the curvature perturbation created during warm
inflation is given by
\begin{eqnarray}
{\cal R}_{warm}\simeq H\frac{\delta \phi}{\dot{\phi}}
\simeq \frac{r_\Gamma^{3/4}}{r_T^{3/2}}\sim \eta^{-3/4}
A_N^{9/8} \left(\frac{\phi_N}{H}\right)^{3/8}.
\end{eqnarray}
From the equation for the number of e-foldings, it is very natural to
expect $\phi_N^{(i)}/\phi_N^{(c)}\sim 1$ for $N_e \sim
60$.\footnote{Note that for the present model the end of remote
inflation is defined by $\phi_N^{(c)}$.}
We thus find a more concrete estimation of the curvature perturbations
given by
\begin{eqnarray}
{\cal R}_{warm}&\simeq&
A_N^{3/2} \left(\frac{r_{T_c}}{\eta}\right)^{3/2}.
\end{eqnarray}
Here $T_c$ is the critical temperature defined for the remote sector,
which determines the end of inflation.
Imposing the normalization given by ${\cal R}\sim 10^{-5}$,
we find an estimation of the coefficient given by
\begin{equation}
A_N\sim 10^{-3.3} \left(\frac{\eta}{r_{T_c}}\right).
\end{equation}
Considering the $O(H)$ correction from the K\"ahler potential in
supergravity, natural value of the $\eta$-parameter is 
$r_\Gamma >\eta \sim 1$.
Low-scale inflation may lead to $\eta \gg1$, which does not ruin the
warm-inflationary scenario as far as the slow-roll conditions 
for warm inflation are satisfied.

Contrary to the usual scenario of generating curvature perturbations at
the end of inflation\cite{End-Modulated, End-multi,
 End-multi-mat}, here a model is considered in which 
the inhomogeneities may be created by the spatial
inhomogeneities of the critical temperature ($\delta
T_c$).\footnote{Note that the creation of the curvature perturbations at
the end of inflation is not general for usual sneutrino
inflation, in which inflation is the chaotic-type.
Instead, there would be an inhomogeneous preheating\cite{IH-PR} at the
end of chaotic-type inflation, which
may cause significant genation of the curvature perturbations with
non-Gaussian character.}
The sources of the inhomogeneities are light fields (moduli), which
determines the value of $g$ and $m_\sigma$. 
From the definition of the critical temperature, we find
\begin{equation}
\frac{\delta T_c}{T_c}=\frac{\delta m_\sigma}{m_\sigma}-
\frac{\delta g}{g}.
\end{equation}
To find the curvature perturbations we must solve 
\begin{equation}
\delta N_{end} \sim H \frac{\delta T_c}{\dot{T}}
\end{equation}
at the end of inflation.
In the present model, $\dot{T}$ is determined by the inflaton motion
$\dot{\phi}_N$ as
\begin{equation}
\dot{T} \sim \frac{M_N^4}{A_N HT^3}\dot{\phi} 
\sim \frac{\eta^2}{A_N r_{T_c}^{3}}\dot{\phi}.
\end{equation}
Therefore, the curvature perturbation created at the inhomogeneous
phase transition is estimated as
\begin{eqnarray}
{\cal R}_{IP}\sim H \frac{\delta T_c}{\dot{T}}
&\sim& A_N r_{T_c}^4 \eta^{-2}\left(\frac{\delta m_\sigma}{m_\sigma}-
\frac{\delta g}{g}\right)\frac{H^2}{\dot{\phi}}\nonumber\\
&\sim& A_N^2 \left(\frac{r_{T_c}^4}{\eta^3}\right)
\left(\frac{\delta m_\sigma}{m_\sigma}-
\frac{\delta g}{g}\right).
\end{eqnarray}
The ratio between ${\cal R}_{warm}$ and ${\cal R}_{IP}$ is given by
\begin{equation}
r_{I/W}\equiv \frac{{\cal R}_{IP}}{{\cal R}_{warm}}
\sim \eta^{-3/2}A_N^{1/2}r_{T_c}^{5/2}
\left(\frac{\delta T_c}{T_c}\right).
\end{equation}
The creation of the curvature perturbations by inhomogeneous phase
transition requires ${\cal R}_{warm}<10^{-5}$, which leads to a more
concrete estimation of the ratio; 
\begin{equation}
r_{I/W} \sim 10^{-5} \times A_N^{-1/2}
\left(\frac{\delta T_c}{H}\right).
\end{equation}

\subsection{Spectral index}
Spectral index of the curvature perturbations in warm inflation has been
calculated in  
Ref.\cite{consistency-warm} for general form of the dissipative 
coefficient.
The calculation presented in the reference is straight and useful.
According to the calculation in Ref.\cite{consistency-warm},
the spectral index for hybrid-like inflation with $\Gamma \propto
T^0$ and $V_{\phi_N}\propto \phi_N^{1+\epsilon_N}$ is given by
\begin{equation}
n_s-1 \simeq -\frac{3\eta}{4r_\Gamma}
\left(\frac{1-2\epsilon_N}{1+\epsilon_N}\right),
\end{equation}
which determines the ratio between $\eta$ and $r_\Gamma$ from the
cosmological observations.

To summarize the results, we find the relation between $A_N$, $r_{T_c}$
and $\eta$ from ${\cal R}_{warm}$, and the equation for
$\eta$, $r_\Gamma$ from the spectral index.
There are two equations for the four parameters
$A_N$, $T_c$, $V_0$ and $M_N$, in which $T_c$ and $V_0$ are the
parameters of the remote sector.
Non-Gaussianity is not significant when the curvature perturbations are
generated by the standard mechanism.
Therefore, observation of the non-linear parameter $f_{NL}\gg 1$
indicates that the curvature perturbations are not created at (and just
after\cite{matsuda-warm}) the horizon exit.

\subsection{Leptogenesis}
\label{lepto-sn}
Leptogenesis can proceed through the out-of equilibrium decay of the
sneutrino inflaton $\phi_N$.
Contrary to the usual sneutrino inflation model, reheating in
remote inflation is caused by the remote-sector phase transition.
Therefore, there is a dilution that leads to a suppression factor 
$\epsilon_D \sim M_N^2 (\phi_N^{(e)})^2/V_0$ for the lepton number density.
The lepton asymmetry of the Universe is measured by the lepton to
entropy ratio that is given by
\begin{equation}
\frac{n_L}{s}\simeq |\epsilon_{CP}|\epsilon_D \frac{T_{RH}}{M_N},
\end{equation}
where $\epsilon_{CP}$ is the CP asymmetry generated at the decay of the
sneutrino inflaton and $T_{RH}$ is the reheating temperature.
A typical bound for the CP-asymmetry parameter is given by\cite{cp-asym}
\begin{equation}
|\epsilon_{CP}| \le 2\times 10^{-8}\frac{M_N}{10^{8}GeV},
\end{equation}
which leads to the condition
\begin{equation}
\frac{n_L}{s} \le
 2\times 10^{-8}\frac{T_{RH}}{10^{8}GeV} \epsilon_D.
\end{equation}

As one would easily understand from the typical potential of the thermal
inflation model, the reheating temperature in sneutrino-remote
inflation can be lowered to a very
small scale, even smaller than the electroweak scale. 
Such a low-reheating temperature is very interesting,
but it obviously 
ruins the successful scenario of leptogenesis caused by the sneutrino
inflaton.
Inflation and the creation of the curvature perturbations are
natural in this sneutrino-remote model, which may work even if the
reheating temperature is very low.
This is an interesting result.
However, leptogenesis may not be sufficient if the reheating
temperature is very low.
The simplest way out of this dilemma would be to find a more general
 inflaton candidate in MSSM fields.\footnote{There are many 
models of baryogenesis that may work with low-scale inflation. See for
example Ref.\cite{matsuda-baryon} and the references therein.}
We consider this possibility in section \ref{mssm-inf}.
However, contrary to the sneutrino-remote inflation model that gives
very simple 
results, arguments for MSSM-remote inflation with general inflaton
potential are rather diverse.
Moreover, the form of $\Gamma$ may be different
depending on the model parameters and the choice of the MSSM fields.
Therefore, the  purpose of the next paragraph is to show essential
equations for 
MSSM-remote inflation when the diffusion parameter is given by
the simplest form $\Gamma \sim A_M
\phi_M$.

\subsection{MSSM field as the inflaton}
\label{mssm-inf}
Applying remote inflation scenario to MSSM inflaton field,
one finds immediately that significant difference
 may appear in the inflaton potential and $\epsilon_{CP}$.
The difference in the inflaton potential may lead to the difference in
the slow-roll conditions, the curvature perturbations and the spectral
index, while the difference in 
$\epsilon_{CP}$ may lead to a significant enhancement of the
 baryon number asymmetry for low-reheating temperature.

When the MSSM inflaton ($\phi_M$) has the (quasi-)flat potential, 
the typical form of the effective inflaton potential for $T>T_c$ is
given by 
\begin{eqnarray}
V(\phi_{M})&=& V_0\left\{1+c'_0\ln \frac{\phi_M}{M_*}\right\}
\equiv V_0 +m^4 \lambda_0 \ln \frac{\phi_M}{M_*}\,\,\,\,\, (n\equiv0),
\end{eqnarray}
or considering non-renormalizable terms the effective 
potential may be characterized by
\begin{eqnarray}
V(\phi_{M})&=& V_0\left\{1+c_n\left(\frac{\phi_M}{M_*}\right)^n
\right\} 
\equiv  V_0 +\lambda_n\frac{\phi_M^{n}}{M_*^{n-4}}
\,\,\,\, (n\ne 0),
\end{eqnarray}
where $c'_0$, $c_n$ and $M_*$ are determined by the choice of the MSSM
field, while $V_0$ is determined by the remote-sector potential.
Although the potential with $n<0$ is unlikely for the MSSM fields, 
it may appear in a hidden sector of dynamical supersymmetry breaking.
In this paper we consider implicitly the potential with $n=0$ and
$n\ge 4$.
The potential characterized by $n=2$ 
is the same as the sneutrino-remote model, although the typical
mass-scale is utterly different.
 
For simplicity, we consider only the typical form of the
dissipative coefficient that is given by
\begin{equation}
\label{basic-eq}
\Gamma \equiv A_M \phi_M,
\end{equation}
where the coefficient $A_M$ is determined by the Yukawa couplings.
The slow-roll velocity of the MSSM inflaton for the strongly dissipating
regime ($r_\Gamma >1$) is given by
\begin{eqnarray}
\dot{\phi}_M &\simeq& -\frac{c'_0 V_0}{\Gamma \phi_M}
\simeq-\frac{c'_0}{A_M}\frac{V_0}{\phi_M^2}
\simeq -\frac{m^4\lambda_0}{A_M \phi_M^2}\,\,\,\, (n=0)\\
\dot{\phi}_M &\simeq&-\frac{n  c_n V_0}{\Gamma \phi_M}
\left(\frac{\phi_M}{M_*}\right)^n
\simeq -\frac{n  c_n V_0}{A_M \phi_M^2}
\left(\frac{\phi_M}{M_*}\right)^n\nonumber\\
&\simeq&
 -\frac{n \lambda_n M_*^2}{A_M}\left(\frac{\phi_M}{M_*}\right)^{n-2}
\, (n\ne0).
\end{eqnarray}

Again, $T$-dependence of $\Gamma$ leads to a condition for the
radiation\cite{warm-inflation-original}
\begin{equation}
c\equiv \left|\frac{d\ln \Gamma}{d\ln T}\right|<4,
\end{equation}
which is satisfied {\bf except} for $1\le n\le
2$\cite{consistency-warm}.

In addition to the standard slow-roll conditions for the inflaton motion, 
slow-roll approximation for warm inflation requires slow-variation of
$\Gamma$, which leads to the additional slow-roll parameter given
by\cite{warm-inflation-original} 
\begin{eqnarray}
\beta &\equiv&
 M_p^2 \left(\frac{\Gamma_{\phi_N}V_{\phi_N}}{\Gamma V}\right)
\simeq \frac{c_0'M_p^2}{ \phi_M^2}
\sim \frac{m^4\lambda_0}{H^2 \phi_M^2}
 < 1+r_\Gamma \,\,\,\,\,\, (n=0),\\
\beta &\equiv& M_p^2 
\left(\frac{\Gamma_{\phi_N}V_{\phi_N}}{\Gamma V}\right)
\simeq \frac{n c_n M_p^2 \phi_M^{n-2}}{M_*^n} 
\sim \frac{n \lambda_n \phi_M^{n-2}}{H^2 M_*^{n-4}}  < 1+r_\Gamma
 \,\,\,\,\,\, (n>2),
\end{eqnarray}
which lead to non-trivial conditions for the inflaton field;
\begin{eqnarray}
\phi_M &>&
 M_p \sqrt{\frac{c_0'}{(1+r_\Gamma)}}\sim 
\frac{m^2\sqrt{\lambda_0}}{H \sqrt{1+r_\Gamma}}\equiv \phi_M^{\beta}
\,\,\,\,(n=0)\\
\phi_M &<& \left(\frac{(1+r_\Gamma)M_*^n}{n c_n M_p^2}
\right)^{\frac{1}{n-2}}
\sim \left(\frac{(1+r_\Gamma)H^2 M_*^{n-4}}{n \lambda_n}
\right)^{\frac{1}{n-2}}
\equiv \phi_M^{\beta}\,\,\,\,(n>2)
\end{eqnarray}
The temperature during inflation is given by
\begin{eqnarray}
T^4 
&\sim& \frac{1}{H\pi^2 g_*} \frac{(c_0')^2 V_0^2}{A_M \phi_M^3}
\sim
\frac{m^8\lambda_0^2}{H\pi^2 g_* A_M \phi_M^3}\,\, (n =0),\\
T^4 &\sim&
\frac{1}{H\pi^2 g_*} \frac{n^2  c_n^2 V_0^2}{A_M \phi_M^3}
\left(\frac{\phi_M}{M_*}\right)^{2n}
\sim
M^4_* \frac{n^2 \lambda_n^2}{\pi^2 g_* A_M}
\left(\frac{M_*}{H}\right)
\left(\frac{\phi_M}{M_*}\right)^{2n-3}\,\, (n>2). 
\end{eqnarray}
Remote inflation is possible during $T>T_c$, which defines 
the critical point given by
\begin{eqnarray}
\phi_M<\phi_M^{(c)}  
&\equiv& \left(\frac{c_0' V_0^2}{H\pi^2 g_*A_M T_c^4}
\right)^{1/3}\nonumber\\
&&\sim  \left(\frac{m^8 \lambda_0^2}{H\pi^2 g_*A_M T_c^4}
\right)^{1/3}
\,\,\, (n = 0),\\
\phi_M>\phi_M^{(c)}  
&\equiv& \left(\frac{H\pi^2 g_*A_M T_c^4 M_*^{2n}}{n^2  c_n^2 V_0^2}
\right)^{\frac{1}{2n-3}}\nonumber\\
&&\sim 
\left(\frac{H\pi^2 g_*A_M T_c^4 M_*^{2n-8}}{n^2  \lambda_n^2}
\right)^{\frac{1}{2n-3}}
\,\,\, (n>2).
\end{eqnarray}
Here $T_c$ is determined by the remote-sector potential.

In addition to these conditions,
the conventional slow-roll conditions are the important conditions that
determine the cosmological quantities created during inflation.
The condition related to the $\epsilon$-parameter is broken when
$\epsilon >1+r_\Gamma$, which leads to the conditions for the slow-roll
inflation given by
\begin{eqnarray}
\phi_M >\phi_M^{\epsilon} &\equiv& M_p\frac{c_0'}{\sqrt{1+r_\Gamma}}\sim
M_p\frac{\lambda_0}{\sqrt{1+r_\Gamma}}\left(\frac{m^4}{V_0}
\right) \,\,\,\,\,(n=0)\\
\phi_M <\phi_M^{\epsilon} &\equiv&  
M_* \left(\frac{\sqrt{1+r_\Gamma}}{n c_n}
\frac{M_*}{M_p}\right)^{\frac{1}{n-1}}
\sim  \left(\frac{\sqrt{1+r_\Gamma} V_0M_*^{n-4}}{n \lambda_n M_p}.
\right)^{\frac{1}{n-1}}(n>2)
\end{eqnarray}
For $n >2$ and $\lambda_n>0$, slow-roll inflation 
is possible during 
$\phi_M^{(c)}<\phi_M<Min\{\phi_M^{\epsilon},\phi_M^\beta\}$,
while for $n=0$ and $\lambda_0>0$, we find that the condition is given by
$Max\{\phi_M^{\epsilon},\phi_M^\beta\}<\phi_M<\phi_M^{(c)}$.
Remote inflation is not successful if these conditions are not satisfied. 
To understand the end-boundary of remote inflation,
it would be useful to consider the difference between $n=0$ and $n>2$.
In contrast with $n >2$, inflation with
the potential characterized by $n=0$ may terminate with the 
violation of the slow-roll condition.
Namely, for the positive coefficient $\lambda_0>0$, inflaton motion 
of the $n=0$ potential is roll-in, and the slow-roll condition is
violated at $\phi_M=Max\{\phi_M^\epsilon, \phi_M^\beta\}$.
The temperature is increasing during inflation, which means that there
is no phase transition during inflation.
However, for the negative coefficient $\lambda_0<0$, inflaton motion is
roll-out, and the inflation ends with the phase transition at 
$\phi_M=\phi_M^{(c)}$.
The temperature is decreasing during inflation.

To understand clearly these arguments, it would be useful to consider
a simple and concrete example in which some parameters are fixed at 
typical values.
To show an explicit example, we find for the simplest case with 
$n=6$;
\begin{equation}
T^4\simeq \frac{\lambda_6^2 \phi_M^9}{H A_M M_*^4},
\end{equation}
which leads to the critical point defined by the phase transition in the
remote sector;
\begin{equation}
\phi_M^{(c)}\simeq  \left(
A_M HT_c^4 M_*^4 \lambda_6^{-2}\right)^{1/9},
\end{equation}
and the critical points from the slow-roll conditions;
\begin{eqnarray}
\phi_M^\epsilon &\simeq& \left(\frac{\sqrt{1+r_\Gamma}V_0 M_*^2}
{\lambda_6 M_p}\right)^{1/5}\\
\phi_M^\beta &\simeq&
\left(\frac{(1+r_\Gamma)H^2 M_*^2}
{\lambda_6 }\right)^{1/4}.
\end{eqnarray}

Obviously, remote inflation is possible for MSSM inflaton fields,
during $\phi_M^{(c)}<\phi_M < Min\{\phi_M^{\epsilon},\phi_M^\beta\}$.
Using $\phi_M^{(c)}$ for the end of remote inflation,
and assuming that $\phi_M^{(i)} \sim \phi_M^{(c)}$
for $N\sim 60$,
we find the curvature perturbations given by
\begin{equation}
{\cal R}_{MSSM}\simeq \frac{r_\Gamma^{3/4}}{r_T^{3/2}}
\sim r_{T_c}^{-7/6}
A_M^{5/6}\lambda_6^{-1/6}\left(\frac{M_*}{H}\right)^{1/3}.
\end{equation}

An intuitive argument would be useful for understanding
the discrepancy 
between the usual MSSM inflation and MSSM-remote
inflation.
Standard MSSM inflation\cite{MSSM-inflation-0} 
realized with tiny Hubble parameter requires tiny 
slow-roll parameter for the creation of the curvature perturbations.
Namely, from the normalization of the curvature perturbations and the
standard formula
\begin{equation}
{\cal R}\sim H\frac{\delta \phi}{\dot{\phi}},
\end{equation}
it is found that the slow-roll parameter 
$\epsilon \sim 10^{-22}(H/10^2 GeV)^2$ is required
to create the cosmological perturbations.\footnote{
Here the creation of the curvature perturbations at the horizon exit 
is discussed.
Curvatons or inhomogeneous preheating may relax the
fine-tuning problem in low-scale inflation scenario
\cite{IH-PR-low, matsuda_curvaton}. } 
Since the condition required for the creation of the curvature
perturbations ($\epsilon < 10^{-22}$) 
is much more severe than the slow-roll
condition $\epsilon<1$, there is a dilemma that $\epsilon$ at the
creation of the curvature perturbations must be far different from
the value at the end of inflation.
Namely, rapid variation of $\epsilon$ is required for the last
e-foldings $N_e\sim 60$, while the potential must be very flat at the
horizon exit.
In contrast with supercooled inflation, strong dissipation in warm
inflation may cause strong damping of the inflaton motion, even if the
Hubble-induced mass leads to $\eta\sim 1$.
Moreover, since the end of remote inflation is not defined by $\epsilon
\sim 1$, 
there is no dilemma related to the variation of $\epsilon$.
If the dissipation rate is given by $\Gamma \sim A\phi_M$,
it enhances the curvature perturbation by the huge factor 
$\Gamma /H \sim r_\Gamma\sim A\phi_M/H\gg 1$.
This factor may cure the fine-tuning problem of the supercooled 
MSSM inflation model.

\subsection{Baryogenesis for MSSM inflaton}
Compared with sneutrino-remote inflation,
a significant discrepancy may appear in the estimation of the baryon
number asymmetry, which is caused by the magnitude of $\epsilon_{CP}$
and the mass of the inflaton field.
A rough estimation of the baryon asymmetry of the Universe 
measured by the baryon to entropy ratio is given by
\begin{equation}
\frac{n_B}{s}\sim |\epsilon_{CP}|\epsilon_D \frac{T_{RH}}{m_{soft}},
\end{equation}
where $T_{RH}$ can be as small as the scale of the
nucleosynthesis\cite{thermal-inflation}. 
The MSSM direction may have either baryon or lepton number.
For high $T_R$, leptogenesis may work with active sphalerons.
The baryon number is conserved if $T_R$ is very low, where there is no
active sphaleron after reheating.

\section{Conclusions}
In this paper we applied a new scenario of hybrid-like inflation
to sneutrino and MSSM inflaton fields.
In remote inflation, radiation raised continuously by a dissipating
inflaton 
field keeps symmetry restoration in a remote sector, and the
false-vacuum energy of the remote sector dominates the energy density 
 during inflation. 
The situation is similar to hybrid inflation, but the hybrid-type
potential is not required in this scenario.
Remote inflation ends when the temperature reaches the critical
 temperature, or when the slow-roll condition is violated.
Using the scenario of remote inflation, we considered sneutrino and MSSM
fields as the inflaton candidate, and found essential conditions for the
inflaton potential and the remote sector.
The results show that remote inflation may naturally 
work with sneutrino or MSSM fields.
More details of the scenario (i.e., more details of the required
parameter space for the curvature perturbations and baryogenesis)
must be studied using numerical calculations.

\section{Acknowledgment}
We wish to thank K.Shima for encouragement, and our colleagues at
Tokyo University for their kind hospitality.

\appendix
\section{Thermal Inflation and reheating in the remote sector}
A generic supersymmetric gauge theory will have a large number of
directions which is flat before supersymmetry breaking.
The minimal supersymmetric standard model (MSSM) will have the flat
directions, but it does not lead to $V_0\ne 0$ during remote inflation.
We thus considered for the remote sector a flat direction (flaton) which
is not the MSSM flat direction but can be thermalized during remote
inflation.\footnote{Note that the remote field must be distinguished
from the inflaton field. The remote field plays the role of the trigger
field that appears in the conventional hybrid inflation.}
The most obvious example of this kind has already been 
discussed for thermal inflation\cite{thermal-inflation}. 
A field with a nonzero VEV is defined either by a Higgs field (which is
not necessarily the standard-model Higgs or the conventional-GUT Higgs
field) or a gauge singlet.
The discussions for the original work of thermal inflation were focused
on the latter case, but it was also commented that in some GUT models a
Higgs field may play the role.
In particular, gauge-mediated models of supersymmetry breaking has a
hidden sector which interacts with the standard-model fields with
renormalizable interactions.
The crucial difference between remote-sector inflation and thermal
inflation appears in the source of the radiation.
In thermal inflation, there is no source of the radiation during
inflation, while in remote inflation dissipation in the inflaton sector 
sources radiation.

With regard to reheating in the remote sector, our scenario is very
similar to thermal inflation, which is {\bf not similar to the reheating
mechanism in MSSM inflation}.\footnote{Note that the oscillation after
thermal inflation does not lead to preheating.
Some details are discussed for MSSM fields in
Ref.\cite{Mazumdar-PR-MSSM}. See also Ref.\cite{Mazumdar-reviewer}. } 
Note that in MSSM-remote inflation, MSSM field appears as the inflaton
in the inflaton sector, however reheating is caused by the
remote-trigger field in the remote sector, which can be identified with 
the flaton in the conventional thermal inflation scenario.
In the remote inflation scenario, reheating is not due to the decay of the
MSSM inflaton, but induced by the remote-trigger field in the remote
sector. We think there is no confusion on this point, since 
the scenario is comparable to the well-known scenario of hybrid inflation.

A flaton particle corresponding to oscillations around a VEV of $M_0$
will couple only weakly to particles with mass less than $M_0$.
The simplest choice of the decay rate is $\Gamma \sim m^3/M_0^2$, where
$m$ is the flaton mass near the VEV.
For example, an effective interaction $\lambda |\sigma|^2||\chi|^2$ 
between the remote field $\sigma$ and a spin-zero particle $\chi$ 
with the mass $m_\chi$ will lead to the decay
rate\cite{thermal-inflation} 
\begin{equation}
\Gamma \simeq \frac{\lambda^2}{8\pi}\left(\frac{M}{m}\right)^2
m\sqrt{1-\frac{4m_\chi^2}{m^2}}.
\end{equation}
Maximizing the decay rate with the obvious condition for the decay
channel\footnote{This condition must not be applied to the interaction
with thermal background during remote inflation.}
\begin{equation}
\label{dec-cond}
\lambda \le \frac{1}{2}\frac{m_\chi^2}{M^2},
\end{equation}
it is found that $\Gamma \le 10^{-4} m^3/M^2$\cite{thermal-inflation}.
Note that there are many choices for the remote sector, which are highly
model-dependent.

The discussions for the flaton reheat temperature in the remote sector
is precisely the same as what has been discussed for thermal inflation.
The typical example that has been discussed in the original work of
thermal inflation leads to the reheat temperature
\begin{equation}
T_D \sim \left(\frac{10^{11}GeV}{M_0}\right)
\left(\frac{m}{300GeV}\right)^{3/2}GeV.
\end{equation}
Of course, there are many choices for the flaton field that lead to
many different values of the reheat temperature.
It is not important to show the catalog of the flaton candidates
in this appendix, since the purpose of this paper is to show a
significant possibility of MSSM-remote inflation.

\section{On thermal corrections to the MSSM inflaton}
If part of the standard-model gauge group (or the extended gauge group
in some extended models of the standard model) remains unbroken by the 
VEV of flat fields, the effective mass of the associated gauge fields
(and gauginos) will remain massless.
The background thermal bath with a temperature $T$ will affect the flat
direction dynamics.
A detailed study of such thermal corrections has been presented in
Ref.\cite{Mazumdar-reviewer} with regard to the scenario in which MSSM
directions are identified with curvatons.
For the MSSM-remote inflation scenario, the situation is
precisely the same as what happens in the conventional warm-inflation
scenario.
Namely, for the MSSM-inflaton VEV $\phi$ that induces a mass $y\phi$
to the field that couples to the MSSM-inflaton, the effective potential
that arises from the thermal background are
\begin{eqnarray}
\Delta V_{th} &\sim& y^2 T^2 |\phi|^2, \,\,\,\,\, (y\phi\le T)\nonumber\\
\Delta V_{th} &\sim& \pm \alpha T^4 \ln \left(|\phi|^2\right),
 \,\,\,\,\, (y\phi > T)
\end{eqnarray}
where $\alpha$ is a gauge fine structure constant.
The obvious condition, which arises for any kind of warm inflation
scenario, is given by $y\phi \gg T$.
Of course, this condition must be satisfied during MSSM-remote
inflation, although it frequently leads to a trivial condition that can
be disregarded compared with other
conditions\cite{warm-inflation-original}.

\end{document}